# Understanding before verifying: Claim normalization for automated citation verification

Yifan He[1,2], Mengjia Wu[2], Siming Deng[3], and Yi Zhang[2]

[1] College of Humanities, Donghua University, Shanghai, China

[2] Australian Artificial Intelligence Institute, Faculty of Engineering and Information Technology, University of Technology Sydney, Sydney, New South Wales, Australia

[3] Institute for Future Initiatives, The University of Tokyo, Tokyo, Japan

**Correspondence**

Yi Zhang

Email: yi.zhang@uts.edu.au

**Other author email addresses**

Yifan He: 1222039@mail.dhu.edu.cn

Mengjia Wu: mengjia.wu@uts.edu.au

Siming Deng: simingdeng@foxmail.com

# Understanding before verifying: Claim normalization for automated citation verification

## Abstract

Citation accuracy has been studied for decades because of its importance to research reliability. Content-level citation verification assesses the reliability of scholarly claims. Recent work adopts a two-stage retrieval–classification framework inherited from fact-checking. However, this design overlooks the complexity of the raw citing claim and introduces three issues into the verification system, namely scope mismatch, perspective mismatch, and proposition entanglement. These issues increase the difficulty of retrieval and classification, thereby limiting model performance. Motivated by this gap, we propose claim normalization, which applies three rewriting strategies to the raw citing claim before retrieval and classification, allowing each downstream model to perform a single, well-defined task. Building on this method, we develop Claim-Normalized Citation Verification (CNCV), a new three-stage framework consisting of claim normalization, evidence retrieval with grounding, and citation classification. We evaluate CNCV across 18 classifiers using a factorial experiment on human-annotated citation instances. Compared with the prior two-stage framework, CNCV improves macro F1 by an average of 12% for encoders and 10% for generative LLMs, driven by improved evidence quality, the dominant factor identified in our experiments. Evidence retrieved from automatically normalized claims yields downstream classification performance statistically equivalent to that obtained with manually annotated evidence.

## 1 INTRODUCTION

Citation verification asks whether a citing statement faithfully represents the source it cites (Pavlovic et al., 2021), and citations that fail to do so are commonly termed miscitations (Simmering et al., 2025), which arise at two layers (Genzinger & Wills, 2017). At the bibliographic layer, researchers have primarily focused on fabricated references (Hicks, 2021; Luo et al., 2013). At the content layer, an equally important but often overlooked question is whether valid references are faithfully represented in the citing text (Baethge & Jergas, 2025). Because underlying inaccuracies can be concealed behind plausible citation claims and legitimate references, such miscitations are often more difficult to detect (Sarol et al., 2024). When such miscitations occur and remain undetected, they may contribute to the persistence and propagation of erroneous knowledge (Ioannidis, 2018; Montenegro et al., 2021).

A growing body of empirical research indicates that a substantial share of citations do not faithfully represent the cited source (Kurlansky et al., 2025; Wakeling et al., 2025). A well-known example is the New England Journal of Medicine letter by Porter and Jick (1980), later widely cited as evidence of low opioid addiction risk (Leung et al., 2017). The accumulation of such cases (Cumberledge et al., 2023; Greenberg, 2009) has raised growing concern in the academic community, yet their detection requires substantial manual effort, prompting the development of automated citation verification systems. Current systems (Liu et al., 2024; Sarol et al., 2024) adopt a two-stage framework inherited from fact-checking (Thorne et al., 2018), which consists of a retrieval stage in which a retriever uses the raw citing claim to select evidence from the cited paper, followed by a classification stage in which a classifier assigns a label based on the claim and retrieved evidence.

However, this design does not account for the complexity of raw citing claims, ultimately reducing verification accuracy. A raw citing claim may contain information that the cited source is not expected to support, be framed from a perspective different from that of the source, and combine multiple propositions. When such a claim is used as a retrieval query (Haan, 2025), these characteristics give rise to scope mismatch, perspective mismatch, and proposition entanglement, respectively. As a result, the retriever may return irrelevant or misaligned evidence, or evidence that supports only some of the propositions. The classifier may therefore lack the evidence needed to make an accurate judgment. The same characteristics also complicate classification when the raw citing claim is used as the anchor for evaluating the retrieved evidence (Liu et al., 2024). The classifier may assess content beyond the scope of the citation, infer an unstated perspective without guidance, or assign unequal weight to entangled propositions. Each of these problems increases the risk of an incorrect verdict. These issues lower the accuracy of the system's judgments but pose no obstacle to human reviewers, since an experienced reviewer readily understands what a citation conveys, knows how to gather the relevant evidence, and can judge it correctly. This explains the large gap reported in a prior study between classifiers given human-annotated claims and evidence and those given the raw claim with automatically retrieved evidence (Sarol et al., 2024). Motivated by this insight, we propose claim normalization, which applies three consecutive rewriting strategies to raw claims before the retrieval and classification stages. The method is designed to emulate the “understand-then-verify” process followed by human reviewers, thereby addressing the three issues and improving the accuracy of automated citation verification.

With claim normalization as its core method, we develop Claim-Normalized Citation Verification (CNCV), a new three-stage framework consisting of claim normalization, evidence retrieval with grounding, and citation classification. Beyond claim normalization, CNCV differs from prior two-stage methods in its retrieval stage. It employs an LLM-based retriever rather than a conventional BM25-based retrieval and cross-encoder reranking pipeline because the LLM can better exploit the semantic and structural information introduced by claim normalization. Within the retrieval stage, we apply evidence grounding to correct occasional substitutions or merged fragments in the LLM output by mapping the retrieved evidence back to the source text through strict textual similarity matching. To evaluate CNCV, we construct a dataset of 1,444 citation instances with human-normalized claims, human-selected gold evidence, and human-assigned labels. This dataset addresses a limitation of existing resources, which rely on automatically generated negative samples (Alvarez et al., 2024) or LLM-generated annotations (Haan, 2025). Each instance also includes the original citing claim, its citation marker, and a set of candidate evidence sentences. Following the severity distinction established in citation-audit research, we distinguish between major and minor miscitation errors (Montenegro et al., 2021; Wakeling et al., 2025). We then evaluate CNCV across 18 commonly used classifiers, demonstrating its effectiveness through ablation experiments and investigating its underlying mechanism through factorial analysis.

Compared with the prior two-stage framework, CNCV improves macro F1 across all 18 classifiers, with average gains of 12% for encoders and 10% for generative LLMs. Factorial analysis shows that evidence quality has a substantially greater effect on classification performance than claim type, with only a small interaction effect between the two factors. The results show that classification performance does not improve continuously with evidence quality. Instead, they suggest that evidence quality must reach a certain threshold for classification models to perform the task effectively, with this threshold varying across task settings.

## 2 RELATED WORK

### 2.1 Automated citation verification

Compared with the decades-long tradition of manual citation auditing (De Lacey et al., 1985; Wakeling et al., 2025), automated citation verification has a short history, and both its conception and its implementation arose in the biomedical domain. Kilicoglu (2018) first proposed using natural language processing (NLP) for citation verification in biomedicine, a proposal grounded in traditional machine learning, since pretrained Transformer models had not yet become standard in NLP. This conception was not implemented for several years. In the interval, research on the broader topic of fact-checking had advanced more rapidly, establishing a two-stage framework that first retrieves evidence and then classifies the claim (Thorne et al., 2018; Wadden et al., 2022). Because citation verification is structurally aligned with fact-checking, Sarol et al. (2024) applied this two-stage framework directly to a biomedical citation dataset, realizing Kilicoglu's conception. After reviewing prior studies on citation verification and conducting small-scale experiments, we find that the two-stage framework is not fully suited to this task and therefore propose CNCV.

### 2.2 Claim normalization

Although claim normalization is designed specifically for citation verification, it builds on a broader principle that has proven effective across multiple tasks, namely reformulating input text to better suit the downstream task. Sundriyal et al. (2023) rewrite a noisy social media post into a normalized claim, thereby providing a clear target for fact-checking. Choi et al. (2021) rewrite a sentence that depends on its context into one that reads on its own, so that it can be interpreted and verified in isolation. Min et al. (2023) break a passage into separate elementary statements, so that each can be checked individually. Ma et al. (2023) rewrite questions as search queries that are better aligned with the retriever, thereby improving both evidence retrieval and downstream question answering. The success of these approaches across different settings supports the use of text reformulation as the underlying principle of claim normalization.

## 3 METHODOLOGY

### 3.1 Definition and task formulation

This study addresses content-level citation verification, which assesses whether a claim in a citing paper is faithfully supported by the cited source associated with a given citation marker. The system is intended as a tool that flags potentially problematic citations for further review by domain experts. As illustrated in Figure 1, the task is defined for one citation marker at a time. The verification system takes the claim and its target marker from the citing paper, together with a candidate evidence sentence set spanning the full text of the cited paper, and assigns one of three labels: Accurate, Minor error, or Major error.

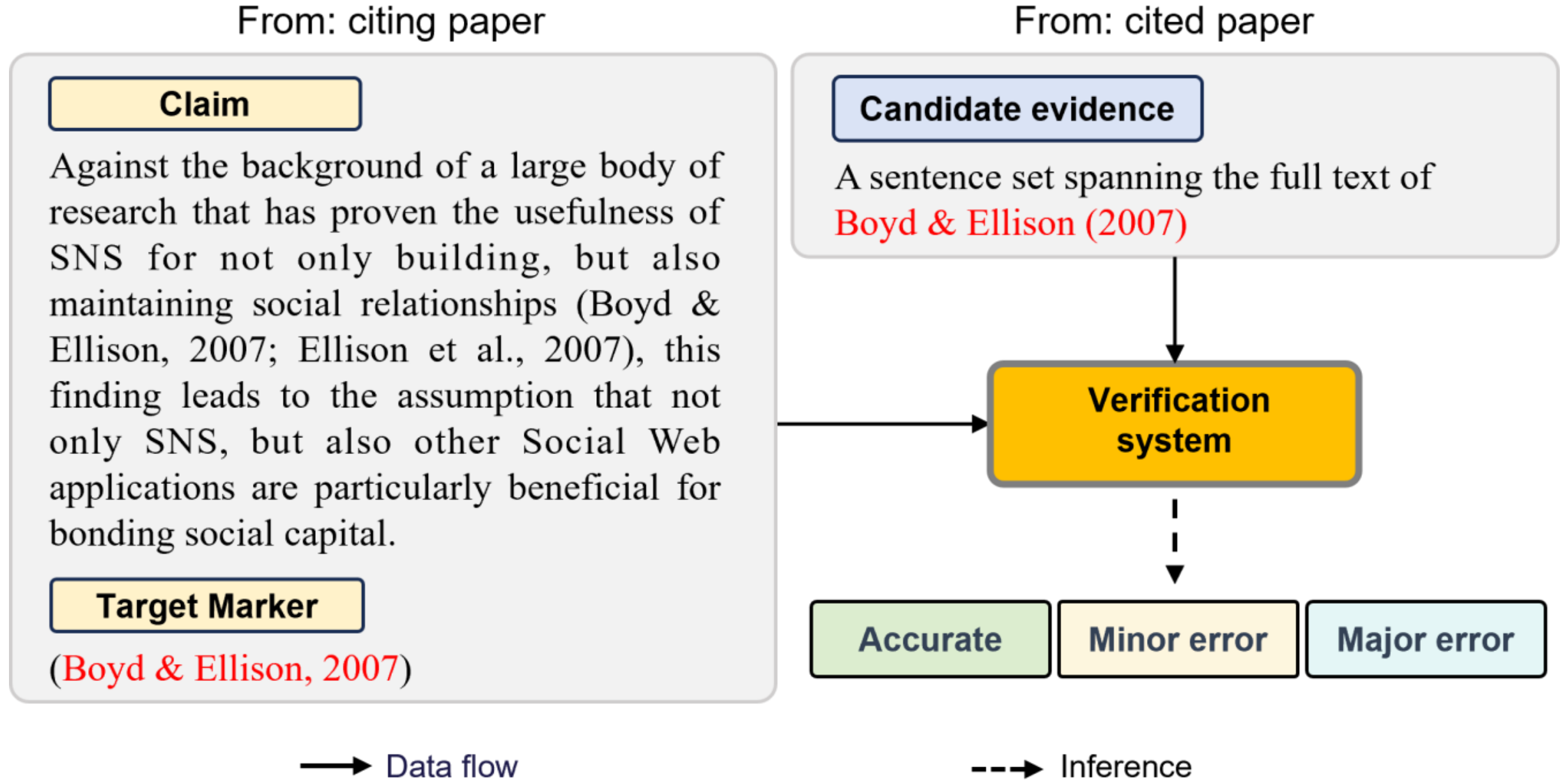

**Figure 1.** Task formulation for content-level citation verification.

The definitions of the three labels and their associated categories are provided in Table 1. We grade miscitations by severity (Montenegro et al., 2021; Wakeling et al., 2025), and adopt the citation classification framework of Sarol et al. (2024) with modifications tailored to our task formulation. This yields six fine-grained labels, each corresponding to a distinct error mechanism, defined in Table 1 (representative cases are detailed in Supplemental Material A). Because a single instance can in principle correspond to more than one label, the labels follow the priority order, and each instance is assigned the highest-priority applicable label, so that the most serious error is identified first. The first four categories cover the cases where the source fails to support the claim, and are graded as major errors. Partially supported means that, with none of those major errors present, the claim nonetheless overreaches that support, which is graded a minor error.

**Table 1.** Definitions and assessment mappings of the six annotation labels.

| Priority | Category | Definition | Label |
|---|---|---|---|
| 1 | Unrelated | The cited paper is unrelated to the topic of the claim and provides neither support nor refutation. | Major error |
| 2 | Peripheral | The cited paper addresses a related topic, but the proposition expressed in the claim is not stated or discussed in the cited paper. | |
| 3 | Misreported | The cited paper discusses the proposition, but specific values or interpretations are inaccurately reported. | |
| 4 | Contradicted | The cited paper explicitly states the opposite conclusion or direction of the claim. | |
| 5 | Partially Supported | The claim has some basis in the cited paper but exceeds the actual support, including over-generalization, out-of-scope extrapolation, or overstated strength. At least one proposition is clearly supported. | Minor error |
| 6 | Accurate | The claim fully and accurately reflects the corresponding statement in the cited paper, with no deviation in scope or strength. | Accurate |

The six fine-grained categories are used only for annotation, whereas all classification experiments use the corresponding three-class scheme. This design directly reflects the objective of the present task, which is to determine whether a miscitation is present and, if so, how severe it is. It also avoids requiring classifiers to distinguish among fine-grained error mechanisms that are not essential to this objective, a choice that accords with the general insight from multiclass learning theory that the size and structure of the label space can affect learning complexity (Brukhim et al., 2022).

### 3.2 Data construction and annotation

We collected all open-access articles indexed under the Information Science & Library Science category in Web of Science and ranked the citing–cited article pairs in descending order based on the citation counts of the cited articles. From the top 2,000 candidate pairs, we manually retrieved 957 usable pairs. Because a single citing–cited article pair may contain multiple in-text citation occurrences, these pairs yielded 1,444 citation instances. The resulting dataset and its accompanying documentation are publicly available (He et al., 2026), as described in the Data Availability Statement.

The dataset was annotated by three experts with backgrounds in information science. The annotation guidelines were developed through multiple rounds of pilot annotation, with revisions after each round to clarify boundary cases. Inter-annotator agreement was substantial under the Landis and Koch (1977) interpretation, with Fleiss' $\kappa = 0.748$ at the six-class granularity and 0.752 at the three-class granularity. The finalized guidelines, the full annotation procedure, and additional agreement metrics are reported in Supplemental Material A. The final dataset is split into 1,154 training, 146 validation, and 144 test samples, with the validation set drawn by stratified sampling across labels. Figure 2 reports the three-class distribution across splits. The observed miscitation rate falls within the range reported in prior citation accuracy studies (Wakeling et al., 2025).

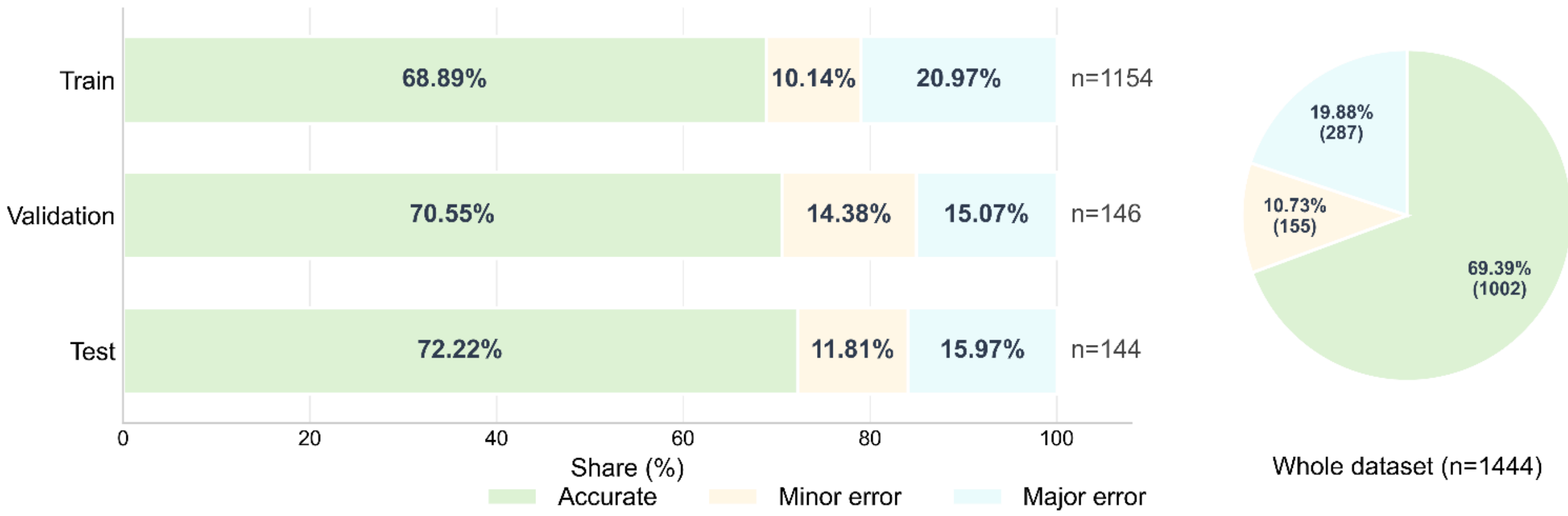

*Note. Counts and within-split percentages are reported. The six-class distribution is provided in Supplemental Material A.*

**Figure 2.** Three-class label distribution across dataset splits.

### 3.3 CNCV

Figure 3 provides an overview of CNCV. First, a prompted LLM (GPT-5-mini) sequentially applies three strategies to rewrite the raw citing claim into a normalized claim. Next, an LLM retriever (GPT-5-mini) uses the normalized claim to select relevant evidence from the candidate evidence set, and a grounding procedure maps the selected evidence back to the corresponding source sentences, producing Evidence (L-norm). Finally, an encoder-based or LLM-based classifier pairs the normalized claim with L-norm and assigns one of three labels.

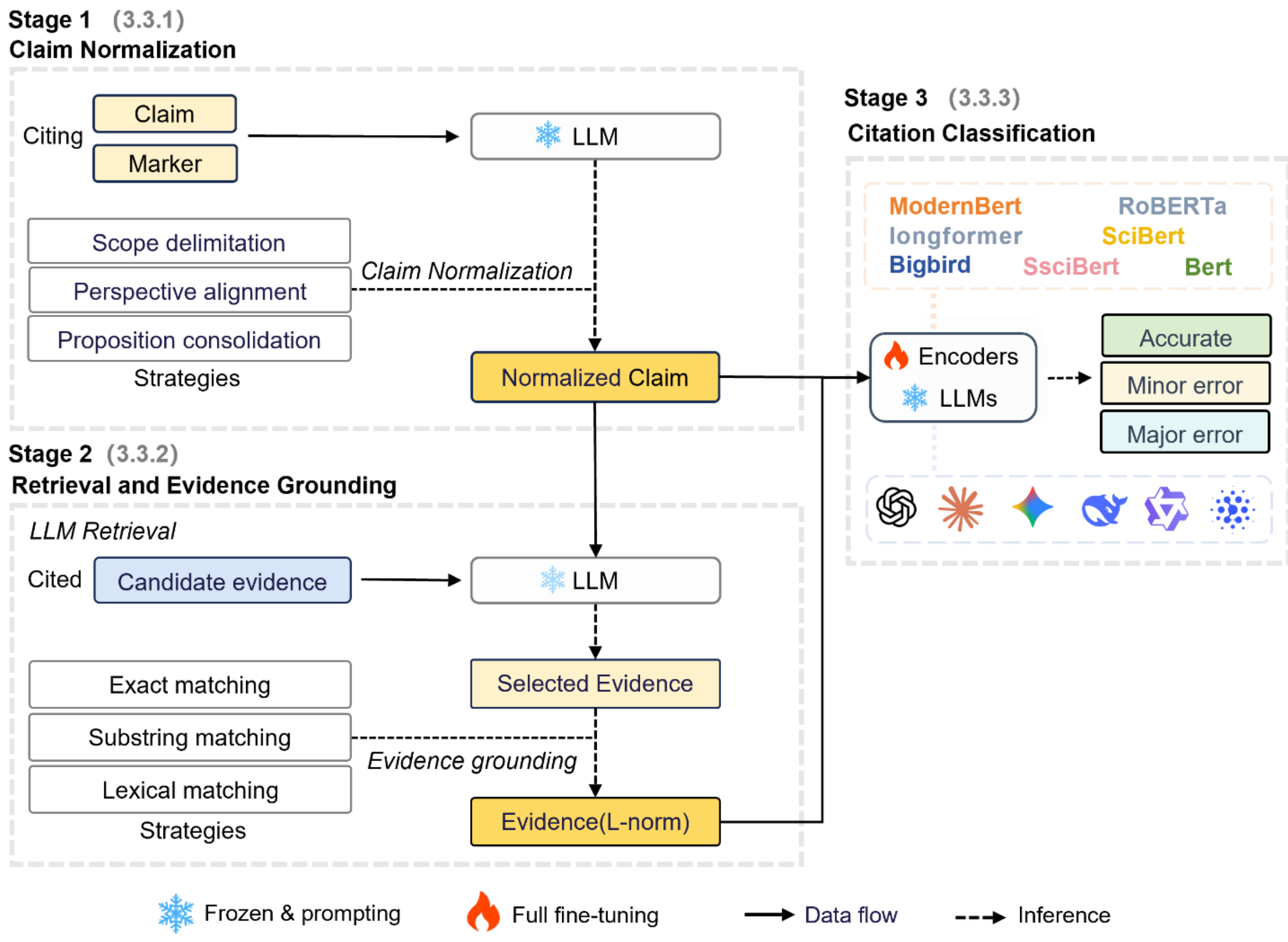

**Figure 3.** Overview of CNCV.

3.3.1 Claim normalization

Figure 4 illustrates claim normalization through an example. The three strategies delimit the portion of the claim that the cited paper is expected to support, align the claim's perspective with that of the paper, and clarify its propositional structure, thereby improving the performance of downstream models in evidence retrieval and citation classification. The order is fixed because each strategy operates on the output of the previous one.

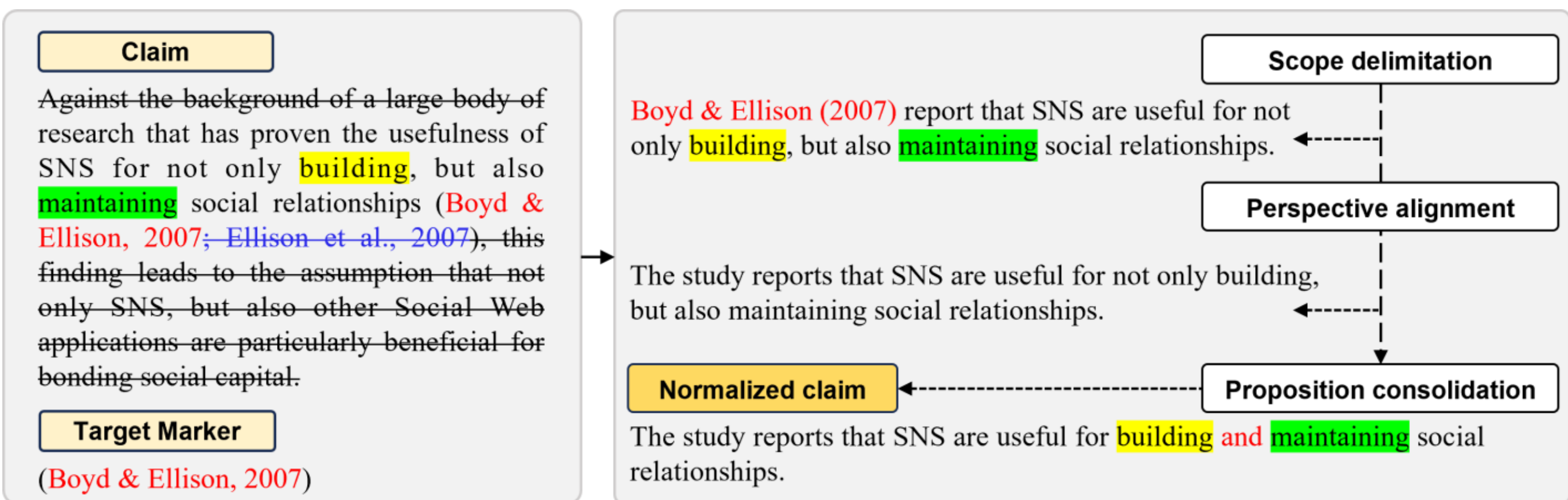

**Figure 4.** Illustration of the three claim normalization strategies.

Scope delimitation keeps only the span supported by the target marker. By default, this span consists of the clause or clauses immediately before the marker, but earlier text is included when necessary to recover scope or referents. The step also separates citing-author additions, such as examples, comparisons, applications, or generalizations, from the marker-supported content. It further removes motivations, implications, or extensions not supported by the target marker to avoid introducing misleading information into retrieval and classification.

Perspective alignment rewrites the delimited span so that it begins with "The study" and uses the cited study as the grammatical subject. This uniform form aligns the claim more directly with the cited paper than do varied author–year expressions, such as "Boyd and Ellison (2007) report that...". This strategy is motivated by our observation that models sometimes fail to associate an author–year expression with the cited paper, even when explicitly prompted to do so, resulting in faithful claims being misclassified as errors.

Proposition consolidation expresses all supported propositions in a single normalized sentence, explicitly linked by "and" when more than one proposition is present. This step preserves every proposition supported by the marker while making the propositional structure visible, so that the model can retrieve evidence and judge the claim against all supported content rather than overweighting only part of an entangled claim.

Table 2 makes explicit the effects that each unresolved issue may have on retrieval, where the claim serves as the query, and on classification, where the claim serves as the input. The full prompt, API configuration and input–output format used for automatic claim normalization are provided in Supplemental Material B.

**Table 2.** Effects of the three issues on retrieval and classification.

| Issue | Retrieval (claim as query) | Classification (claim as input) |
|---|---|---|
| Scope mismatch | The claim's out-of-scope content disrupts retrieval, leaving the classifier without the key evidence | The whole claim leads the classifier to judge content beyond the span the marker supports, so claims faithful within that span may be misclassified as errors |
| Perspective mismatch | The perspective gap between claim and evidence disrupts retrieval, so the sentences that would support the claim are never retrieved | The classifier must bridge the shift in grammatical perspective and recognize the claim's legitimate generalization or instantiation, otherwise accurate claims may be misclassified as errors |
| Proposition entanglement | The claim's compound propositions, used as the query, make the retrieved evidence cover them unevenly, leaving some propositions without evidence to verify them against | Without grasping the claim's compound propositions, the classifier treats them as an undifferentiated whole and weights them unevenly, yielding a verdict that reflects only part of the claim |

### 3.3.2 Retrieval and evidence grounding

The candidate evidence is constructed by segmenting the full text of each cited paper into sentences using spaCy (Honnibal et al., 2020). The retriever receives the normalized claim and the candidate sentences and returns up to ten sentences that support or bear on the claim as retrieved evidence—a cap chosen because the human-annotated evidence contains an average of 7.27 sentences per claim and already contains some redundancy. Implementation details, including the LLM prompt, are provided in Supplemental Material C.

We use an LLM for retrieval rather than the conventional BM25-based retrieval and cross-encoder reranking pipeline because the three approaches differ in how fully they can exploit the structure introduced by claim normalization. BM25 is primarily a lexical ranking function based on term frequency, inverse document frequency, and document length normalization (Robertson & Zaragoza, 2009). Although normalization may still improve lexical matching by removing noise and emphasizing salient terms, BM25 cannot directly model the semantic and structural relations introduced by the rewritten claim. A cross-encoder jointly encodes each claim and candidate sentence and applies self-attention across their tokens to produce a relevance score, allowing it to capture semantic correspondence beyond lexical overlap (Nogueira & Cho, 2019). However, although it reranks the recalled candidates according to these scores, it evaluates each claim–candidate pair independently. By contrast, an LLM with sufficient context capacity and reasoning ability can process the complete claim together with all candidate sentences in a single context, enabling joint comparison and reasoning over the candidate set. After surveying the models available as of April 2026, we selected GPT-5-mini (OpenAI, 2025a) as the best-performing option within our cost constraints.

As a generative model, the LLM may return evidence sentences that paraphrase or shorten the original source sentences. Following Rashkin et al. (2023), we apply evidence grounding as a post-processing step within the retrieval stage to anchor each LLM-generated evidence sentence to a verifiable source sentence. Specifically, each generated sentence is aligned with one of the candidate source sentences using the three-step matching strategy defined in Table 3. Fewer than 0.1% of the generated sentences fail all three matching steps and are therefore discarded as unmatched.

**Table 3.** Operations and trigger conditions of the three-step matching strategy.

| Step | Operation | Trigger condition |
|---|---|---|
| Exact matching | Compare normalized strings of LLM output and candidate sentence; accept if identical | Default |
| Substring matching | Accept if the LLM output is a substring of a candidate sentence, or vice versa; take the longest match | Exact matching fails |
| Lexical matching | Score candidates by token-level Jaccard similarity and recall over content words; accept the top candidate if it satisfies the threshold conditions* | Substring matching fails |

*Note.* *Lexical matching accepts a candidate only when all five thresholds are met: (i) Recall ≥ 0.70; (ii) Jaccard ≥ 0.25; (iii) intersection size ≥ 4 content tokens; (iv) length ratio ∈ [0.5, 5.0]; (v) score margin ≥ 0.01 over the runner-up. Full details are provided in Supplemental Material D.*

### 3.3.3 Classification

The verdict is produced by a classifier given a claim and a set of evidence sentences. Following recent dual-family evaluation designs (Wu, Sivertsen, et al., 2025; Wu et al., 2026), we evaluate encoder-based and LLM-based classifiers separately. We fine-tune nine widely used encoder-based classifiers, running five random seeds per model to control training stochasticity. Nine representative LLM classifiers from six major providers are evaluated under the zero-shot learning setting. Each encoder classifier receives a structured string combining the normalized claim and the L-norm as input, whereas each LLM classifier additionally receives a structured prompt. The full configuration is provided in Supplemental Material E.

**Table 4.** Encoder and LLM classifiers.

| Type | Models |
|---|---|
| Encoder | BERT-large (Devlin et al., 2019), RoBERTa-large (Liu et al., 2019), SciBERT (Beltagy et al., 2019), SSCI-BERT (Shen et al., 2023), BigBird-base, BigBird-large (Zaheer et al., 2020), Longformer-base, Longformer-large (Beltagy et al., 2020), ModernBERT-large (Warner et al., 2025) |
| LLM | GPT-5-mini (OpenAI, 2025a), GPT-5-nano (OpenAI, 2025b), Claude-Haiku-4.5 (Anthropic, 2025), Gemini-3-Flash, Gemini-3.1-Flash-Lite (Google, 2025), DeepSeek-V4-Pro (DeepSeek, 2026), GLM-5.1 (Z.AI, 2026), Qwen-3.6-Plus, Qwen-3.6-Flash (Alibaba Cloud, 2026) |

## 3.4 Experimental design and analysis

Our experiments aim to answer two main questions. First, how much does CNCV improve classification performance over the two-stage framework? Second, what mechanism underlies the observed improvement? We address both questions using a fully factorial design that crosses claim type with evidence type, yielding 18 experimental conditions (Figure 5). We describe the two factors below and then explain how this design addresses both questions. In this study, each experimental configuration is defined by pairing claim type (X1) with evidence type (X2). We refer to the configuration using human-normalized claims (C-Gold) and human-annotated evidence (E-Gold) as the gold-input baseline. Classification performance for each model under each configuration is evaluated against gold-standard, human-annotated three-class labels.

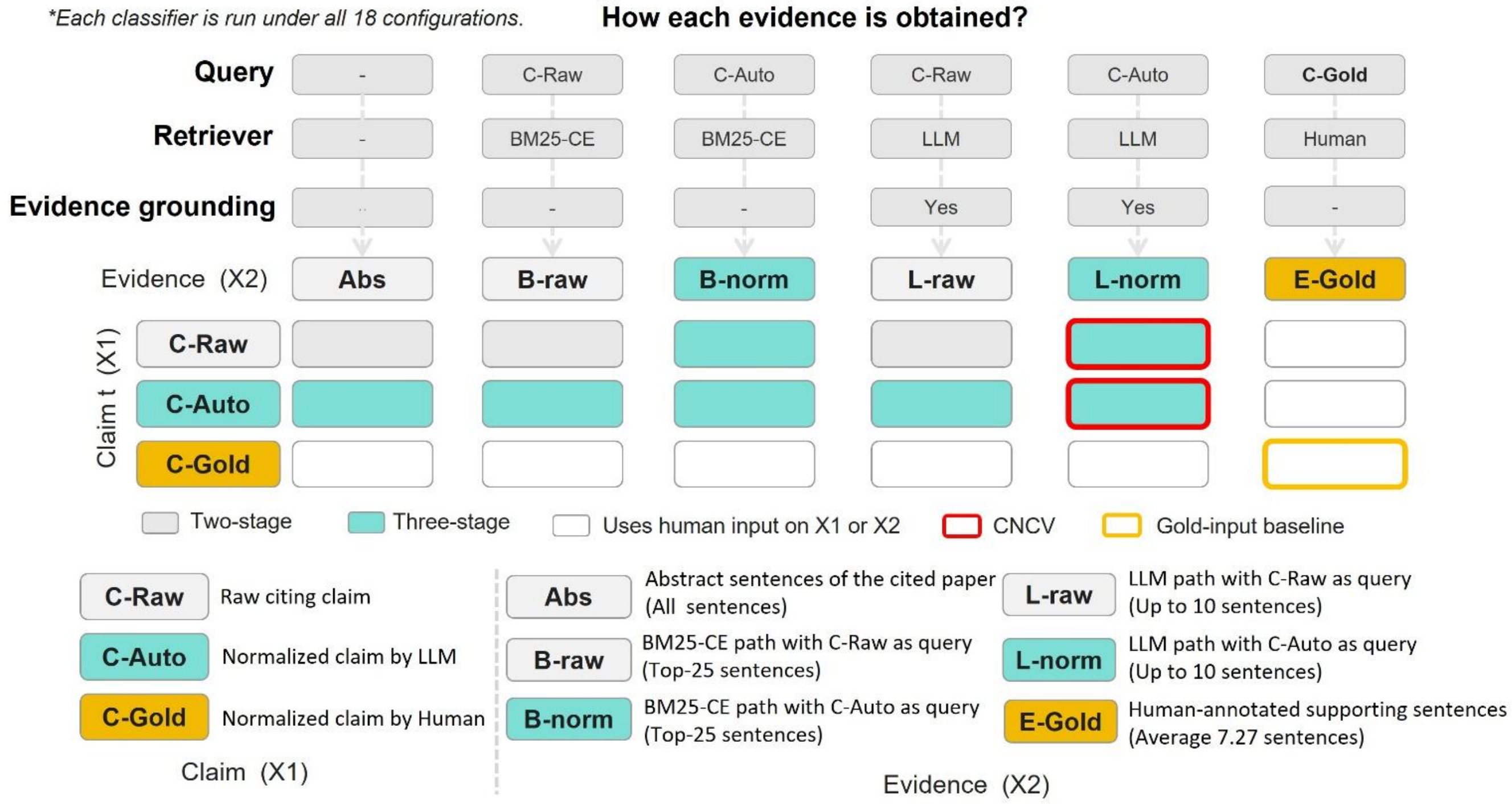

**Figure 5.** Factorial and ablation experiment design.

The first factor is claim type and has three levels. These are the raw citing claim (C-Raw), the LLM-normalized claim (C-Auto), and the human-normalized claim (C-Gold). The second factor, evidence type, comprises six levels that differ in the query used, the retrieval method, and whether evidence grounding is applied, as summarized in the top three rows of Figure 5. Abstract sentences (Abs) are extracted directly from the cited paper. Four retrieval-based levels are produced using either BM25 retrieval followed by reranking with the ms-marco-MiniLM-L6-v2 (Sentence Transformers) (BM25-CE) or the LLM retriever, with either C-Raw or C-Auto as the query. The four retrieval-based evidence types follow a consistent naming convention. The prefix indicates the retriever, with B denoting BM25-CE and L denoting LLM. The suffix indicates the query type, with -raw denoting C-Raw and -norm denoting C-Auto. This convention yields B-raw, B-norm, L-raw, and L-norm. The BM25-CE path first retrieves the top 1,000 candidates using BM25, reranks them with the cross-encoder, and retains the top 25. Abs serves as the no-retrieval baseline, whereas E-Gold serves as the human-annotated evidence baseline.

To answer the first question, the automated conditions fall into two groups (Figure 5), the two-stage baseline that applies no normalization and the three-stage configurations that apply it on at least one side. Comparing the two groups shows whether normalization improves on the prior method, and comparing configurations within the three-stage group identifies the best automated CNCV configuration, which is then compared with the gold-input baseline (C-Gold, E-Gold).

Although the factorial and ablation experiments directly compare performance across configurations, descriptive comparisons alone cannot determine whether the observed differences are systematic across classifiers or distinguish the relative contributions of claim type and evidence type. We therefore apply three complementary analyses to the results of these experiments to answer the mechanism question. All inferential analyses are conducted in Pingouin (Vallat, 2018), using macro F1 as the measure of classifier performance.

First, to determine whether the claim side or the evidence side constitutes the primary performance bottleneck, we conduct a two-way repeated-measures ANOVA with claim type (X1) and evidence type (X2) as within-subject factors. Each subject is evaluated under all 18 combinations, with each model–seed pair treated as one subject in the encoder analysis and each LLM classifier treated as one subject in the LLM analysis. The Greenhouse–Geisser correction is applied to all within-subject effects, and generalized eta-squared ($\eta^2 G$) is reported as the effect size (Greenhouse & Geisser, 1959; Olejnik & Algina, 2003). The relative contributions of claim type and evidence type are assessed by comparing their main-effect sizes, while the interaction measures their departure from additivity.

Second, to examine differences among the levels within each factor, we conduct post hoc paired t-tests on subject-level marginal means, with Bonferroni correction and Hedges' g (Dunn, 1961; Lakens, 2013; Student, 1908). Because a nonsignificant difference does not establish equivalence, any level that does not differ significantly from its corresponding gold-input level—C-Gold for claim type or E-Gold for evidence type—is further evaluated using the two one-sided tests (TOST) (Lakens, 2017; Schuirmann, 1987) procedure with a prespecified equivalence margin.

Third, to investigate how evidence quality relates to classifier performance, we evaluate the four evidence types generated by the BM25-CE and LLM retrieval paths—B-raw, B-norm, L-raw, and L-norm—using E-Gold as the reference and macro recall, retrieval macro F1, MRR, and NDCG@10 as the evaluation metrics. We then descriptively compare these retrieval metrics with the corresponding classifier macro F1 scores across the four evidence types to examine whether improvements in retrieval quality coincide with improvements in classification performance.

### 3.5 Metrics

Table 5 lists all metrics. Macro F1 is the primary classification metric, chosen because the dataset is dominated by accurate citations. Accuracy, precision, and recall are also computed for completeness and reported in Supplemental Material F.

**Table 5.** Evaluation metrics.

| Applied to | Metric | Description |
|---|---|---|
| Classifiers | Macro F1 | Mean of the class-wise F1 scores across the three verdict classes |
| | Accuracy | Proportion of correctly classified samples |
| | Precision | Macro-averaged precision across three classes |
| | Recall | Macro-averaged recall across three classes |
| Retrievers | Recall | Proportion of relevant sentences retrieved |
| | MRR | Mean reciprocal rank of first relevant sentence |
| | NDCG@k | Ranking quality of top-k retrieved sentences |
| LLM Classifiers | Cost (USD) | Inference cost per 100 samples in USD |
| | Latency (s) | Mean inference time per sample in seconds |

*Note. Classifier metrics are computed on the test set (n = 144). Retrieval metrics are computed against human-annotated supporting sentences.*

## 4 RESULTS AND DISCUSSION

### 4.1 Effectiveness of CNCV

We divide the automated configurations into the two-stage group without normalization and the three-stage group with it. For each classifier, we compare its best macro F1 score in the two-stage group with that in the three-stage group and separately report its performance under the gold-input baseline (Table 6, Figure 6). Across all classifiers, CNCV improves macro F1 over the best two-stage configuration of prior methods (Sarol et al., 2024). The mean improvement is 12 % for encoders and 10 % for generative LLMs, with corresponding SDs of 3 and 5 %. Every classifier shows an improvement, indicating that the benefit extends across multiple architectures, pretraining domains, and providers. For half of the classifiers, the best three-stage configuration achieves an F1 score equal to or higher than the corresponding gold-input baseline.

**Table 6**. Best two-stage and best three-stage macro F1 for each classifier, with the gold-input baseline.

| Family | Model | Best2 F1 | Best2 cfg | Best3 F1 | Best3 cfg | Delta | Gold-input F1 | Best3 ≥Gold-input |
|---|---|---|---|---|---|---|---|---|
| Encoder | BERT-large | 0.40 | (C-Raw, B-raw) | 0.55 | (C-Auto, L-norm) | 0.15 | 0.55 | Yes |
| | BigBird-base | 0.44 | (C-Raw, Abs) | 0.57 | (C-Auto, L-norm) | 0.13 | 0.57 | Yes |
| | BigBird-large | 0.40 | (C-Raw, L-raw) | 0.55 | (C-Auto, L-norm) | 0.15 | 0.53 | Yes |
| | Longformer-base | 0.46 | (C-Raw, L-raw) | 0.57 | (C-Auto, L-norm) | 0.11 | 0.60 | - |
| | Longformer-large | 0.47 | (C-Raw, L-raw) | 0.60 | (C-Auto, L-norm) | 0.13 | 0.62 | - |
| | ModernBERT-large | 0.45 | (C-Raw, Abs) | 0.58 | (C-Raw, L-norm) | 0.13 | 0.57 | Yes |
| | RoBERTa-large | 0.49 | (C-Raw, Abs) | 0.56 | (C-Auto, L-norm) | 0.07 | 0.59 | - |
| | SSCI-BERT | 0.39 | (C-Raw, Abs) | 0.52 | (C-Auto, L-norm) | 0.13 | 0.52 | Yes |
| | SciBERT | 0.44 | (C-Raw, Abs) | 0.52 | (C-Auto, L-norm) | 0.08 | 0.53 | - |
| | MEAN±SD | 0.44±0.03 | | 0.56±0.03 | | 0.12±0.03 | 0.56±0.03 | 5/9 |
| LLM | Claude-Haiku-4.5 | 0.39 | (C-Raw, L-raw) | 0.47 | (C-Auto, L-norm) | 0.08 | 0.50 | - |
| | DeepSeek-V4-Pro | 0.34 | (C-Raw, B-raw) | 0.36 | (C-Raw, L-norm) | 0.02 | 0.33 | Yes |
| | GLM-5.1 | 0.36 | (C-Raw, L-raw) | 0.44 | (C-Auto, L-norm) | 0.08 | 0.42 | Yes |
| | GPT-5-mini | 0.60 | (C-Raw, L-raw) | 0.69 | (C-Raw, L-norm) | 0.09 | 0.75 | - |
| | GPT-5-nano | 0.55 | (C-Raw, L-raw) | 0.69 | (C-Auto, L-norm) | 0.14 | 0.72 | - |
| | Gemini-3-Flash | 0.59 | (C-Raw, B-raw) | 0.65 | (C-Auto, L-norm) | 0.06 | 0.63 | Yes |
| | Gemini-3.1-Flash-Lite | 0.56 | (C-Raw, L-raw) | 0.67 | (C-Raw, L-norm) | 0.11 | 0.63 | Yes |
| | Qwen-3.6-Flash | 0.38 | (C-Raw, L-raw) | 0.48 | (C-Raw, L-norm) | 0.10 | 0.52 | - |
| | Qwen-3.6-Plus | 0.45 | (C-Raw, L-raw) | 0.63 | (C-Auto, L-norm) | 0.18 | 0.65 | - |
| | MEAN±SD | 0.47±0.11 | | 0.56±0.13 | | 0.10±0.05 | 0.57±0.14 | 4/9 |

*Note. Best2 F1 and Best3 F1 are each classifier's highest macro F1 within the two-stage and three-stage groups, and cfg gives the winning (claim, evidence) configuration. Delta is Best3 F1 minus Best2 F1. Gold-input F1 is the macro F1 under (C-Gold, E-Gold), and Best3 ≥ Gold-input is Yes when Best3 F1 reaches it.*

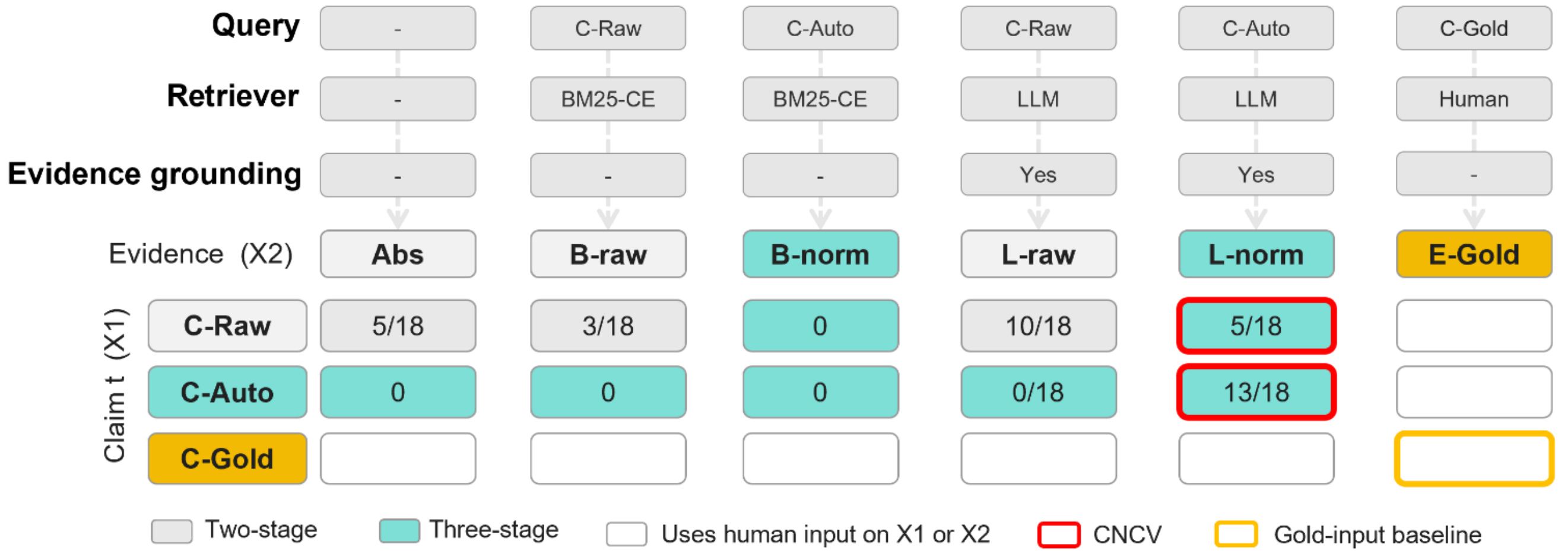

**Figure 6.** Distribution of best two-stage and best three-stage configurations across the 18 classifiers.

L-norm is the winning automated evidence condition for all 18 classifiers, whereas C-Auto is retained as the classification input in 13 of the 18 winning configurations. This asymmetry suggests that normalization contributes more consistently through the retrieval query than through the claim presented to the classifier. This can be attributed to the way normalization removes out-of-scope information, aligns the perspective, and restructures entangled propositions before retrieval, thereby helping the LLM retriever identify evidence that more directly addresses the proposition under verification. Once such evidence is available, some classifiers can still perform well with C-Raw.

### 4.2 Evidence dominance

Table 7 reports the results of a two-way repeated-measures ANOVA of classifier macro F1 for the encoder classifiers, with claim type (X1) and evidence type (X2) as within-subject factors. By crossing the two factors, the design estimates each main effect independently of the other, separating the contributions of the two sides—the claim and the evidence. Both main effects and the interaction are significant under the Greenhouse-Geisser correction, but their effect sizes separate the two sides sharply. By generalized eta-squared (Olejnik & Algina, 2003), evidence type dominates claim type by a ratio of approximately 3.7:1 (0.45 versus 0.12), marking it as the primary determinant of classifier performance, while the interaction, though significant, is small (0.04), so the two sides contribute essentially additively. Consistent with this pattern, combining the best-performing claim level with the best-performing evidence level yields the highest overall performance.

**Table 7.** Two-way RM-ANOVA on classifier macro F1.

| Source | SS | ddof1 | ddof2 | MS | F | p_unc | p_GG_corr | ng2 | eps |
|---|---|---|---|---|---|---|---|---|---|
| X1 | 0.415 | 2 | 88 | 0.208 | 68.227 | *** | *** | 0.122 | 0.997 |
| X2 | 2.447 | 5 | 220 | 0.489 | 110.571 | *** | *** | 0.450 | 0.690 |
| X1 * X2 | 0.112 | 10 | 440 | 0.011 | 4.627 | *** | *** | 0.036 | 0.552 |

*Note. p_unc = uncorrected p-value; p_GG_corr = Greenhouse–Geisser–corrected p-value; ng2 = generalized eta-squared ($\eta^2 G$); eps = Greenhouse–Geisser ε. *** $p < .001$*

A significant main effect does not indicate which of a factor's levels differ, so we resolve each with post-hoc paired t-tests on subject-level means (Tables 8 and 9). On the evidence side, the six types split into two tiers. L-norm and E-Gold form the top tier, each outperforming the other four (Abs, B-raw, B-norm, L-raw) with large effects, while the four do not differ significantly, and a TOST test establishes that L-norm and E-Gold are statistically equivalent. On the claim side, the C-Auto-versus-C-Raw contrast has only a medium effect size (Table 9). In the LLM family, evidence dominates claim even more strongly (a generalized eta-squared ratio of approximately 11:1, compared with 3.7:1 for the encoders), and L-norm again reaches equivalence with E-Gold, but C-Auto and C-Raw no longer differ significantly (Supplemental Material F). One possible explanation is that some of the more capable LLM classifiers can implicitly normalize raw claims during inference, reducing the additional benefit of explicitly providing an LLM-normalized claim (C-Auto).

**Table 8.** Pairwise comparisons among evidence types (X2).

| A | B | mean_A | mean_B | T | p_unc | p_corr_bonf | hedges_g |
|---|---|---|---|---|---|---|---|
| L-norm[a] | E-Gold | 0.551 | 0.550 | 0.274 | 0.785 | 1.000 | 0.042 |
| L-norm | Abs | 0.551 | 0.444 | 10.97 | *** | *** | 2.196 |
| L-norm | L-raw | 0.551 | 0.439 | 17.189 | *** | *** | 2.515 |
| L-norm | B-norm | 0.551 | 0.430 | 15.866 | *** | *** | 3.021 |
| L-norm | B-raw | 0.551 | 0.425 | 16.722 | *** | *** | 3.289 |
| E-Gold | Abs | 0.550 | 0.444 | 10.443 | *** | *** | 2.183 |
| E-Gold | L-raw | 0.550 | 0.439 | 18.783 | *** | *** | 2.504 |
| E-Gold | B-norm | 0.550 | 0.430 | 17.618 | *** | *** | 3.017 |
| E-Gold | B-raw | 0.550 | 0.425 | 17.889 | *** | *** | 3.289 |
| Abs | L-raw | 0.444 | 0.439 | 0.491 | 0.626 | 1.000 | 0.094 |
| Abs | B-norm | 0.444 | 0.430 | 1.599 | 0.117 | 1.000 | 0.269 |
| Abs | B-raw | 0.444 | 0.425 | 2.080 | * | 0.651 | 0.362 |
| L-raw | B-norm | 0.439 | 0.430 | 1.094 | 0.280 | 1.000 | 0.181 |
| L-raw | B-raw | 0.439 | 0.425 | 1.612 | 0.114 | 1.000 | 0.279 |
| B-norm | B-raw | 0.430 | 0.425 | 0.692 | 0.492 | 1.000 | 0.101 |

*Note. Positive g indicates mean A > mean B. [a] For L-norm vs E-Gold, a TOST equivalence test (ε = 0.025 macro F1) confirms statistical equivalence: mean difference +0.001, 90% CI [−0.007, +0.009], p_TOST < .001. * p < .05, ** p < .01, *** p < .001.*

**Table 9.** Pairwise comparisons among claim types (X1).

| A | B | mean_A | mean_B | T | p_unc | p_corr_bonf | hedges_g |
|---|---|---|---|---|---|---|---|
| C-Gold | C-Auto | 0.503 | 0.468 | 7.423 | *** | *** | 0.935 |
| C-Gold | C-Raw | 0.503 | 0.449 | 11.812 | *** | *** | 1.610 |
| C-Auto | C-Raw | 0.468 | 0.449 | 3.909 | *** | *** | 0.581 |

*Note. Positive g indicates mean A > mean B.*

### 4.3 Evidence quality and classifier performance

Table 10 compares the gains in retrieval quality produced by claim normalization along the BM25-CE and LLM retrieval paths. For most automated evidence conditions, improvements in retrieval quality do not produce proportional gains in classifier performance. Macro recall increases from 37.92% for B-raw to 45.43% for B-norm and then to 50.14% for L-raw, yet the three conditions remain statistically indistinguishable in classifier macro F1 (Table 8). Only L-norm, which reaches 64.35% macro recall and performs best on the other retrieval metrics, yields classifier performance equivalent to E-Gold.

**Table 10.** Retrieval quality gains from claim normalization (C-Raw → C-Auto).

| Path | X2 | Macro Recall | Retrieval Macro F1 | MRR | NDCG@10 |
|---|---|---|---|---|---|
| | B-raw | 37.92% | 16.58% | 0.46 | 0.26 |
| BM25-CE | B-norm | 45.43% | 19.68% | 0.57 | 0.33 |
| | Delta | +7.51 pp | +3.10 pp | +0.11 | +0.07 |
| | L-raw | 50.14% | 46.00% | 0.68 | 0.52 |
| LLM | L-norm | 64.35% | 61.67% | 0.81 | 0.67 |
| | Delta | +14.21 pp | +15.67 pp | +0.13 | +0.15 |

This pattern suggests that classifier performance improves only when the retrieved evidence becomes sufficiently complete and well ranked. BM25-CE benefits from the normalized query, but its lexical matching may not fully exploit the revised scope, perspective, and propositional structure. The LLM retriever shows larger gains across all four retrieval metrics, which is consistent with its greater ability to use the semantic changes introduced by claim normalization. Claim normalization therefore contributes most effectively when paired with a retriever capable of using the normalized query.

The four evidence conditions are neither sufficient nor intended to establish a universal recall threshold because both the query representation and the retrieval mechanism vary across conditions. The present experiment therefore cannot identify the precise point at which improvements in retrieval quality translate into gains in classification performance. In practice, rather than targeting a fixed recall threshold that may not be known in advance, system development should aim to improve evidence quality as much as possible. Taken together, the results indicate that the earlier two-stage pipeline is constrained by evidence retrieved from raw claims using conventional retrieval methods. Claim normalization, particularly when paired with LLM-based retrieval, addresses this bottleneck by improving the evidence supplied to the classifier.

### 4.4 Classifier performance

We now take a model-by-model view, examining how individual classifiers perform and how the two families differ. Figure 7 shows the per-model performance surface across all 18 conditions for each encoder and LLM classifier. For the LLMs, it additionally reports the average per-item inference time and the cost per 100 samples. Each panel carries two markers, namely the best overall configuration and the best automated one. Across all 18 models, the best configurations cluster on L-norm and E-Gold evidence, consistent with the results and conclusions of the preceding sections.

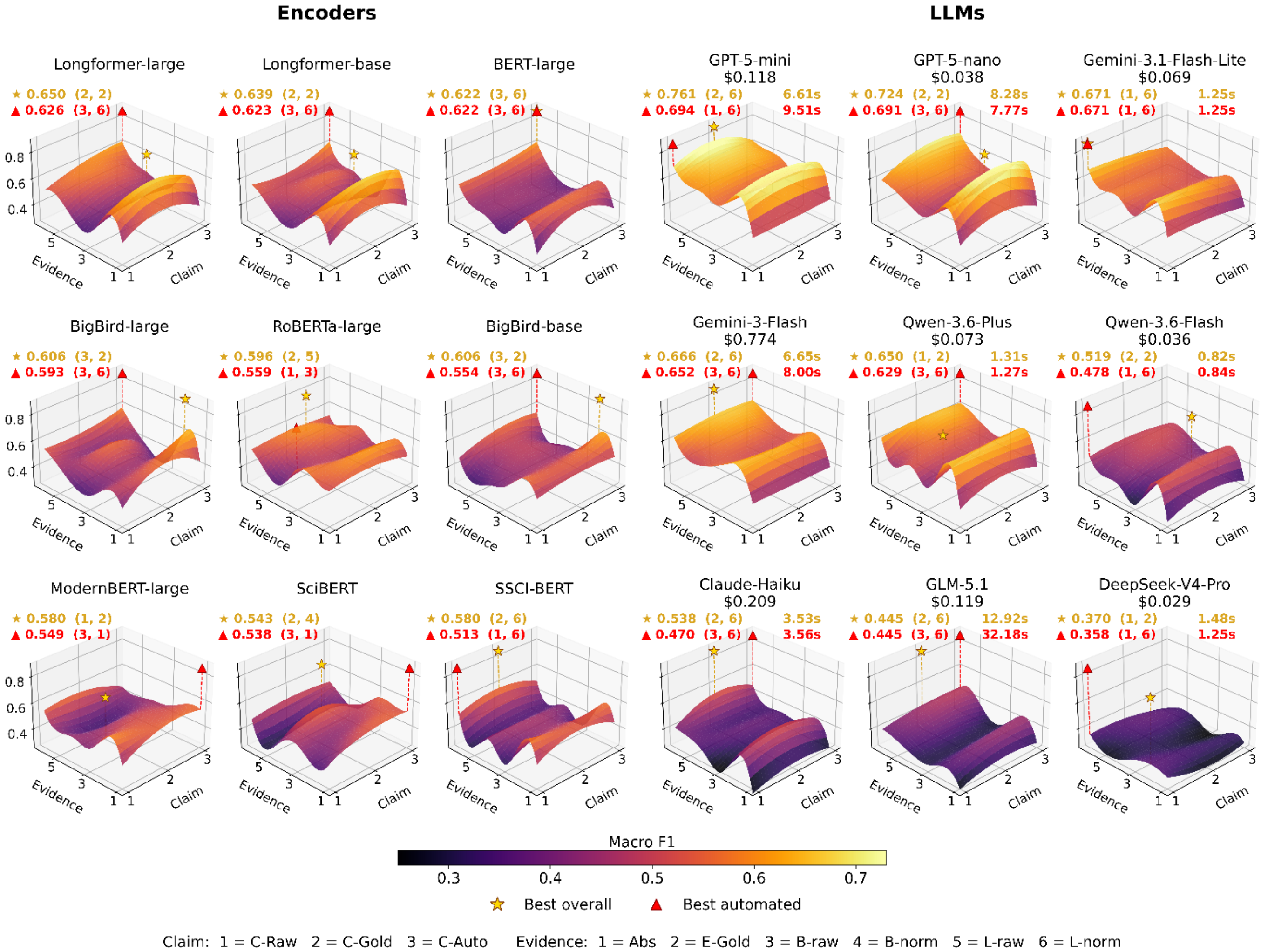

**Figure 7.** Per-model performance landscape across encoders and LLMs.

Five LLMs from three providers exceed the best encoder (Longformer-large at 0.601), while several fall below nearly all fine-tuned encoders. Two caveats apply to the LLM scores. First, GPT-5-mini is used in both the normalization and retrieval stages. When used as a classifier, GPT-5-mini may therefore enjoy a within-family advantage because it receives inputs aligned with its own output style. Accordingly, we do not interpret its performance as an unbiased estimate of the best achievable model performance. This possibility does not affect the conclusion that claim normalization improves performance, since every non-GPT LLM also achieves its best automated F1 with L-norm evidence. Even if part of the gain arises from improved stylistic alignment between the normalized input and downstream classifiers, this would still represent a legitimate benefit of normalization. Second, all LLMs are evaluated zero-shot only, with no few-shot prompting or fine-tuning (Wu, Zhang, et al., 2025); therefore, the reported results should be regarded as conservative estimates.

### 4.5 Cost and practical feasibility

Table 11 reports the end-to-end costs of CNCV for three high-performing classifiers. Across all three, retrieval is the primary cost contributor, costing several times as much as the other two stages combined. This is mainly because the full text of each cited article must be provided as input to the retrieval stage. Consequently, the choice of classifier has only a relatively small effect on the total cost of CNCV, allowing users to select higher-performing classifiers without substantially increasing the overall cost.

**Table 11.** Estimated Cost of Citation Verification

| Classification Method | Claim Normalization (USD) | Retrieval (USD) | Classification (USD) | Total Cost per Citation (USD) | Cost for 10,000 Citations (USD) |
|---|---|---|---|---|---|
| Longformer | 0.001 | 0.01 | — | 0.011 | 110.00 |
| GPT-5 mini | 0.001 | 0.01 | 0.00118 | 0.01218 | 121.80 |
| GPT-5 nano | 0.001 | 0.01 | 0.00038 | 0.01138 | 113.80 |

At the current per-citation cost, CNCV is sufficiently affordable for individual users who verify a limited number of papers. However, the cumulative cost may become substantial when CNCV is deployed at scale by publishers or journal editorial offices. These estimates assume that each citation is processed independently, while actual expenditure may vary with article length, model pricing, and the reuse of retrieved evidence.

For publishers and journals with large-scale verification workloads, costs could be reduced by selectively verifying high-priority citations rather than processing every citation. Priority may be given to citations supporting central conclusions, quantitative findings, methodological decisions, or claims with substantial scientific or practical consequences.

In practice, we recommend using CNCV as an automated screening and early-warning tool. It can be applied to all citing claims within a paper or only a selected subset. Users or domain experts can then prioritize individual claims assigned the “Major error” label, as well as high-risk papers containing multiple such flags, for manual review. The specific review protocol can be tailored to users’ needs.

## 5 CONCLUSIONS

### 5.1 Findings

By improving both the claim and evidence channels, claim normalization substantially increases overall system performance. CNCV improves macro F1 over the best two-stage framework for all 18 classifiers, with average gains of 12% for encoders and 10% for generative LLMs.

Of the two roles played by the raw citing claim, its role as the retrieval query is more consequential. Across both classifier families, evidence type explains substantially more variance in macro F1 than claim type, with this difference being particularly pronounced among LLMs. The results suggest that improvements in retrieval quality translate into better classification only when the retrieved evidence becomes sufficiently informative. The level of retrieval quality required to achieve this sufficiency is not fixed and may vary with dataset construction. Evidence retrieved using automatically normalized claims yields downstream classification performance statistically equivalent to that obtained with manually annotated evidence. These findings identify evidence quality as the primary bottleneck and show that claim normalization delivers most of its gains by improving retrieval rather than merely providing classifiers with a cleaner claim.

The anchor role also contributes, but its effect is smaller than the query role. This smaller effect manifests differently across the two classifier families. For encoders, replacing C-Raw with C-Auto yields a significant gain. For generative LLMs, C-Auto does not significantly outperform C-Raw, yet C-Gold still outperforms C-Auto. A cautious interpretation is that stronger models may both infer part of the needed normalization from raw claims and benefit more from genuinely higher-quality normalized claims. The first tendency would narrow the gap between C-Raw and C-Auto, while the second would preserve the gap between C-Auto and C-Gold. Claim normalization therefore still matters for the anchor role when stronger LLMs are used as classifiers, and the quality gap between automatic and human normalization remains an important target for future improvement.

Several additional findings follow from the analysis. First, the small interaction between claim type and evidence type suggests that the retrieval and classification channels contribute largely additively, so future systems can improve them separately. Second, classifier choice remains consequential. LLMs are not uniformly better than fine-tuned encoders, and their performance is not consistently associated with inference cost. Deployment therefore requires selecting the best-performing classifier within a feasible cost range. Finally, the LLM results should be viewed as conservative, since all LLM classifiers are evaluated zero-shot and may improve with few-shot prompting or fine-tuning.

This work also addresses a content-level gap not covered by current bibliographic-level hallucinated-citation detection tools, which verify whether a cited reference exists and whether its metadata match the bibliographic record but do not assess whether the cited paper actually substantiates the claim it is invoked to support (Sakai et al., 2026).

### 5.2 Implications

The findings establish a clear priority for the design of citation verification systems. Development efforts should focus first on the evidence side, where the main bottleneck lies, with the goal of improving retrieval until it provides evidence that is sufficiently informative for classifier performance to approach the gold-input baseline. Once further improvements in evidence quality yield little additional benefit, the claim-side channel becomes the primary remaining source of performance gains.

### 5.3 Limitations

The findings have two limitations. First, the corpus is drawn from a single domain, Library and Information Science, so generalization to fields with different citation practices remains untested and may alter the observed partitioning of variance. Second, claim normalization and LLM-guided retrieval are implemented using GPT-5-mini only. Although it was selected as the best-performing model within our cost constraints, the observed gains may partly depend on model-specific generation and retrieval behavior, and it remains unclear whether they would generalize to other LLM families. Cross-model replication is therefore an important next step.

### 5.4 Future work

Future work can build on this study in three main directions, all focused on improving claim normalization. First, future studies should compare alternative LLMs and model combinations to identify the best-performing model for claim normalization and determine whether using models from different families for retrieval and classification outperforms using models from the same family. Second, the normalization process itself could be improved through prompt optimization and task-specific post-training, with the goal of narrowing the gap between LLM-generated C-Auto and human-normalized C-Gold. Third, CNCV could be extended beyond the three issues addressed here by identifying additional problems in citing claims and developing corresponding normalization strategies.

**Data Availability Statement**

The data supporting the findings of this study are publicly available at https://github.com/IntelligentBibliometrics/CNCV. Additional supporting information is provided in the Supplemental Materials accompanying this article.

**Generative AI Use Statement**

Generative large language models were used in the proposed framework for claim normalization and evidence retrieval and as zero-shot citation classifiers alongside encoder-based models. Their use is detailed in the Methodology section and Supplemental Materials. Separately, GPT-5.6 and Claude Opus 4.8 were used solely for language editing in parts of the manuscript. All AI-assisted language edits were reviewed and approved by the authors, who take full responsibility for the final manuscript.

**Acknowledgement**

Mengjia Wu was supported by Australian Research Council under Discovery Early Career Researcher Award DE260101493.